\documentclass{epl2} 
\usepackage{epsfig}
\usepackage{amsmath}
\title{Strong pressure-energy correlations in liquids as a configuration space property: Simulations of temperature down jumps and crystallization}
\shorttitle{Strong pressure-energy correlations as a configuration space property}
\author{Thomas B. Schr{\o}der, Ulf R. Pedersen, Nicoletta Gnan, and Jeppe C. Dyre}
\institute{DNRF Centre ``Glass and Time,'' IMFUFA, Department of Sciences, Roskilde University, Postbox 260, DK-4000 Roskilde, Denmark}
\abstract{
Computer simulations recently revealed that several liquids exhibit strong correlations between virial and potential energy equilibrium fluctuations in the NVT ensemble [U. R. Pedersen {\it et al.}, Phys. Rev. Lett. {\bf 100}, 015701 (2008)]. In order to investigate whether these correlations are present also far from equilibrium constant-volume aging following a temperature down jump from equilibrium was simulated for two strongly correlating liquids, an asymmetric dumbbell model and Lewis-Wahnstr{\"o}m OTP, as well as for SPC water that is not strongly correlating. For the two strongly correlating liquids virial and potential energy follow each other closely during the aging towards equilibrium. For SPC water, on the other hand, virial and potential energy vary with little correlation as the system ages towards equilibrium. Further proof that strong pressure-energy correlations express a configuration space property comes from monitoring pressure and energy during the crystallization (reported here for the first time) of supercooled Lewis-Wahnstr{\"o}m OTP at constant temperature.
}

\begin{document}

\maketitle

\section{Introduction}

The subject of thermal equilibrium fluctuations is old and well understood. For large systems equilibrium fluctuations are approximately Gaussian. For any system the ``Gaussian components'' of the fluctuations -- their correlation functions -- determine the system's response to external fields. This is summarized in the noted fluctuation-dissipation theorem known for half a century \cite{lin_resp}, and little new is expected to be learned from studying equilibrium fluctuations

It was recently shown that several model liquids exhibit strong correlations between the thermal equilibrium fluctuations of the configurational parts of pressure and energy at constant volume \cite{ped08a,ped08b,cos08,bai08a}. Recall that the pressure $p$ is a sum of the ideal gas term $Nk_BT/V$ and a term reflecting the interactions, $W/V$, where $W$ is the so-called virial \cite{lin_resp}:

\begin{equation}\label{Wdef}
pV \,=\,
Nk_BT+W\,.
\end{equation}
Temperature is defined via the kinetic energy \cite{lin_resp}, so the ideal-gas pressure term is a function of the particle momenta. The instantaneous virial $W$ is a function of the particle positions, $W=W({\bf r}_1,..., {\bf r}_N)$. In the same way, of course, the energy $E$ is a sum of the kinetic energy and the potential energy $U$. If $\Delta U$ is the instantaneous potential energy minus its average and $\Delta W$ the same for the virial, at any given state point the $WU$ correlation coefficient $R$ is defined by (where sharp brackets denote equilibrium NVT ensemble averages)

\begin{equation}\label{R}
R \,=\,
\frac{\langle\Delta W\Delta U\rangle_{\text {\rm\tiny NVT}}}
{\sqrt{\langle(\Delta W)^2\rangle_{\text {\rm\tiny NVT}}\langle(\Delta U)^2\rangle_{\text {\rm\tiny NVT}}}}\,.
\end{equation}
Strongly correlating liquids by definition have $R>0.9$. For simplicity we shall occasionally talk about ``strong pressure-energy correlations'' although it is only the configurational parts of pressure and energy -- virial and potential energy -- that correlate strongly. 

Strongly correlating liquids include \cite{ped08a,ped08b,cos08,bai08a} the standard Lennard-Jones (LJ) liquid (and crystal), the Kob-Andersen binary LJ liquid, various binary LJ type mixtures, a dumbbell-type liquid of two different LJ spheres with fixed bond length \cite{dumbbell_simulations}, a system with exponential repulsion, a seven-site united-atom toluene model, the Lewis-Wahnstr{\"o}m ortho-terphenyl (OTP) model, and an attractive square-well binary model. Liquids that are not strongly correlating include water and methanol models \cite{bai08a}. The physical understanding developed in Refs. \cite{bai08a,bai08b} is that strong pressure-energy correlations are a  property of van der Waals liquids and some or all metallic liquids. Liquids with directional bonding like covalent and hydrogen-bonding liquids do not have strong pressure-energy correlations. Likewise, ionic liquids are not expected to be strongly correlating because of the different distance dependence of their short-range repulsions and the Coulomb interactions -- competing interactions spoil the correlations.

Strongly correlating liquids appear to have simpler physics than liquids in general, an observation that has particular significance for the highly viscous phase \cite{gt_rev}. It has been shown that supercritical (experimental) argon is strongly correlating \cite{ped08a,bai08b}, that strongly correlating viscous liquids have all eight frequency-dependent thermoviscoelastic response functions \cite{ell07,chr08} given in terms of just one \cite{ped08b} (are ``single-parameter liquids,'' i.e., have dynamic Prigogine-Defay ratio close to unity \cite{bai08b}), that strongly correlating viscous liquids obey density scaling, i.e., that their relaxation time $\tau$ depends on density $\rho$ and temperature as $\tau\propto F(\rho^\gamma/T)$ \cite{density_scaling_exp}, and that even complex systems like a biomembrane may exhibit significant pressure-energy correlations for their slow degrees of freedom \cite{ped08c}. 

\begin{figure*}
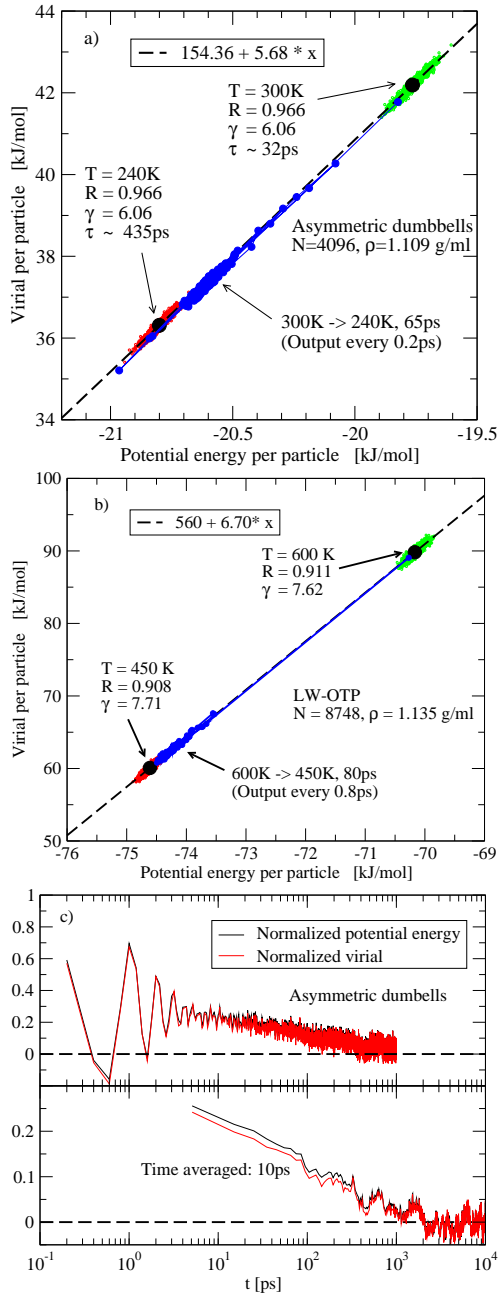

\begin{center}
 \onefigure[width=6.5cm]{fig1a.eps} 
 \onefigure[width=6.5cm]{fig1b.eps}
 \onefigure[width=6.5cm]{fig1c.eps}
\end{center}
\caption{Computer simulations of virial and potential energy during the aging of two strongly correlating liquids following temperature down jumps at constant volume (NVT ensemble) \cite{SimDet}.
(a) The asymmetric dumbbell model at density $\rho = 1.109$ g/ml \cite{dumbbell_simulations}. The liquid was first equilibrated at T=300 K. Here simultaneous values of virial and potential energy are plotted for several times producing the green ellipse, the elongation of which directly shows the strong correlation in equilibrium. Temperature was then changed to T=240 K where the red ellipse marks the equilibrium fluctuations after equilibration. The aging process itself is given by the blue points. These points follow the line defined by the two equilibrium simulations, showing that pressure and energy correlate also out of equilibrium. 
(b) Similar temperature down jump simulation of the Lewis-Wahnstr{\"o}m OTP system \cite{lew94} with quench data plotted every 0.8 ps. The colors have the same meaning as in (a): Green marks the high-temperature equilibrium (T=600 K), red the low-temperature equilibrium (T=450 K), and blue the aging towards equilibrium. -- 
Note that in both (a) and (b) the slope of the dashed line is not precisely the number $\gamma$ of Eq. (\ref{gamma}); this is because the liquids are not perfectly correlating (the line slope is approximately $\langle\Delta U\Delta W\rangle\langle(\Delta U)^2\rangle$ \cite{bai08a}).
(c) Virial and potential energy for the asymmetric dumbbell model as functions of time after the temperature jump of (a); in the lower subfigure data were averaged over 10 ps. Virial and potential energy clearly correlate closely.
}
\label{WU_aging_scl}
\end{figure*}

Whenever equilibrium fluctuations of viral are plotted versus those of the potential energy for a strongly correlating liquid, an elongated ellipse appears \cite{ped08a,cos08,bai08a}. The slope $\gamma$ of this ellipse is given by

\begin{equation}\label{gamma}
\gamma \,=\,
\sqrt{
\frac{\langle(\Delta W)^2 \rangle_{\text {\rm\tiny NVT}} }{\langle(\Delta U)^2 \rangle_{\text {\rm\tiny NVT}}}
}\,.
\end{equation}
This quantity (that is weakly state-point dependent) is the number entering into the above-mentioned density scaling relation \cite{sch08a,sch08b,cos09}. This means that for strongly correlating liquids knowledge of equilibrium fluctuations at one state point provides a prediction about how the relaxation time varies with density and temperature.

What causes strong $WU$ correlations? A hint comes from the fact that an inverse power-law pair potential, $v(r)\propto r^{-n}$ where $r$ is the distance between two particles \cite{ipl}, implies perfect $WU$ correlation ($R=1$) \cite{ped08a,bai08b}. In this case $\gamma=n/3$. In simulations of the standard LJ liquid we found $\gamma\cong 6$ which corresponds to $n\cong 18$. Although this may seem puzzling at first sight given the expression defining the LJ potential, $v_{LJ}(r)=4\epsilon[(r/\sigma)^{-12}-(r/\sigma)^{-6}]$, if one fits the repulsive part of the LJ potential by an inverse power law, an exponent $n\cong 18$ in fact is required \cite{ped08a,bai08b,ben03}. This is because the attractive $r^{-6}$ term makes the repulsion steeper than the bare repulsive $r^{-12}$ term would imply. 

Reference \cite{bai08b} gave a thorough discussion of the correlations with a focus on the LJ case, including also a treatment of the classical crystal where $0.99<R<1$ at low temperature (an anharmonic effect that survives the $T\rightarrow 0$ limit). According to Ref. \cite{bai08b} the $r$-dependent effective exponent $n$ which controls the correlation is not simply that coming from fitting the repulsive part of the potential, but rather $n^{(2)}(r)\equiv -2-rv'''(r)/v''(r)$. This number is approximately 18 around the LJ minimum; in fact the LJ potential may here be fitted very well with an ``extended'' inverse power-law potential \cite{bai08b}, $v_{\rm LJ}(r)\cong A r^{-n}+B+Cr$ where $n\cong 18$; for this particular potential of course $n^{(2)}(r)=n$. At constant volume the linear term contributes little to the viral and potential-energy fluctuations. Thus almost correct Boltzmann probability factors are arrived at by using the inverse power-law approximation. The equation of state, however, is poorly represented by the inverse power law approximation because this potential has no attractive part \cite{bai08b}.

\begin{figure}
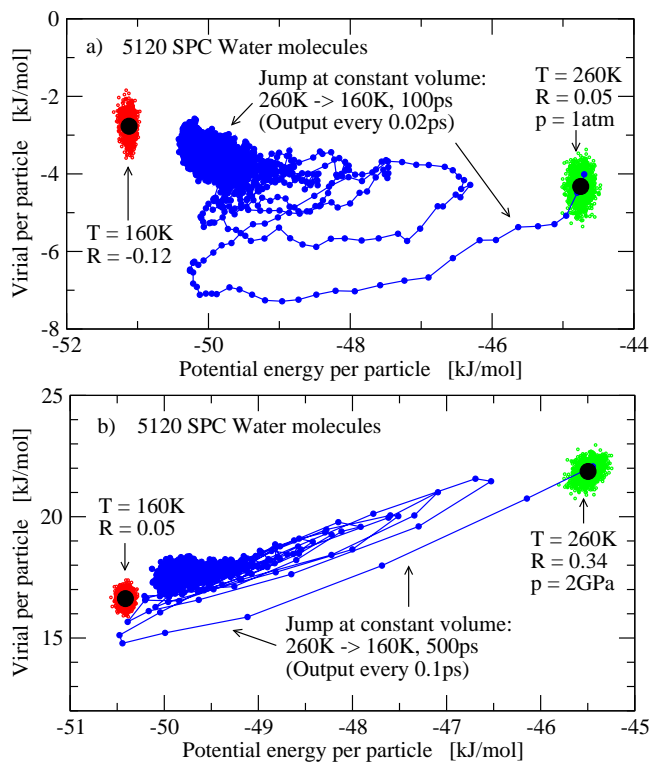

\begin{center}
 \onefigure[width=8.5cm]{fig2a.eps} 
 \onefigure[width=8.5cm]{fig2b.eps}
\end{center}
\caption{Virial versus potential energy after a temperature down jump at constant volume applied to SPC water that is not strongly correlating (colors as in Fig. 1). 
(a) SPC water at 1 atm equilibrated at T=260K, subsequently subjected to a temperature jump down to T=160 K. The data do not follow any line, but are pretty much all over the plane.
(b) Same simulation at 2GPa where there are somewhat stronger $WU$ correlations. The picture is the same, just slightly less confusing.
}
\label{WU_aging_water}
\end{figure}

\section{Temperature down jumps of three model liquids}
If the potential at constant volume to a good approximation may be replaced by an inverse power-law pair potential, strong $WU$ correlations should be present also when the system is not in thermal equilibrium. That is, according to the inverse power-law explanation strong $WU$ correlations characterize all configurations at a given volume. This can be tested in computer simulations where a non-equilibrium situation is easily created by, for instance, abruptly changing temperature for a well equilibrated system.

Figure \ref{WU_aging_scl}(a) shows the results for a temperature down jump at constant volume, starting and ending in equilibrium (NVT ensemble). The system studied is the asymmetric dumbbell liquid consisting of two different sized LJ particles glued together by a bond of fixed length \cite{ped08b}, a toy model for toluene. The system was first equilibrated at 300 K. The green ellipse consists of several simultaneous instantaneous values of $U$ and $W$ in equilibrium at T=300 K. The strong $WU$ correlation is revealed by the elongation of the ellipse ($R=0.97$; $\gamma = 6.1$). When the liquid is similarly equilibrated at 240 K, the red blob appears. To test for correlation in an out-of-equilibrium situation we changed temperature abruptly from the 300 K equilibrium to 240 K. The blue points show how virial and potential energy change following the temperature down jump. Clearly, strong $WU$ correlations are present also during the aging towards equilibrium. Figure 1(b) shows the same phenomenon for the Lewis-Wahnstr{\"o}m (LW) ortho-terphenyl (OTP) model that consists of three Lennard-Jones spheres at fixed length and angle with parameters optimized to mimic real OTP  \cite{lew94}. LW OTP is also strongly correlating ($R=0.91$; $\gamma\cong 7.6$). The colors are as in Fig. 1(a): Green gives the high-temperature equilibrium $T=600$ K, red the $T=450$ K equilibrium, and the blue points show the aging towards equilibrium after changing temperature from 600 K to 450 K. The picture is the same as in Fig. 1(a): The blue points follow the dashed line. Thus virial and potential energy correlate strongly also for far-from-equilibrium states. Figure 1(c) plots $W(t)$ and $U(t)$ for the Fig. 1(a) data for the asymmetric dumbbell liquid. $W(t)$ and $U(t)$ follow each other closely both on picosecond time scales and in their slow, overall drift to equilibrium.

What happens when the same simulation scheme is applied to a liquid that is not strongly correlating? An example is SPC water, where the hydrogen bonds are mimicked by Coulomb interactions \cite{ber87} (any model like SPC water with a density maximum cannot have significant pressure-energy correlations \cite{ped08a,bai08a}). Figure 2 shows results of simulations of SPC water at two different densities, (a) corresponding to low pressure and (b) to very high pressure. In the first case virial and potential energy are virtually uncorrelated ($R=0.05$ at T=260 K); in the second case correlations are somewhat stronger ($R=0.34$ at T=260 K) though still weak. As in Fig. 1 green denotes the initial high-temperature equilibrium, red the low-temperature equilibrium, blue the aging towards equilibrium. Clearly, $W$ and $U$ are not closely linked to one another. The same is apparent from Fig. 2(b), but here is better $WU$ correlation, consistent with the fact that $R$ is larger than in Fig. 2(a).

\section{Pressure and energy monitored during crystallization of a supercooled liquid: The Lewis-Wahnstr{\"o}m OTP model}
\begin{figure*}
\begin{center}
 \onefigure[width=6.5cm]{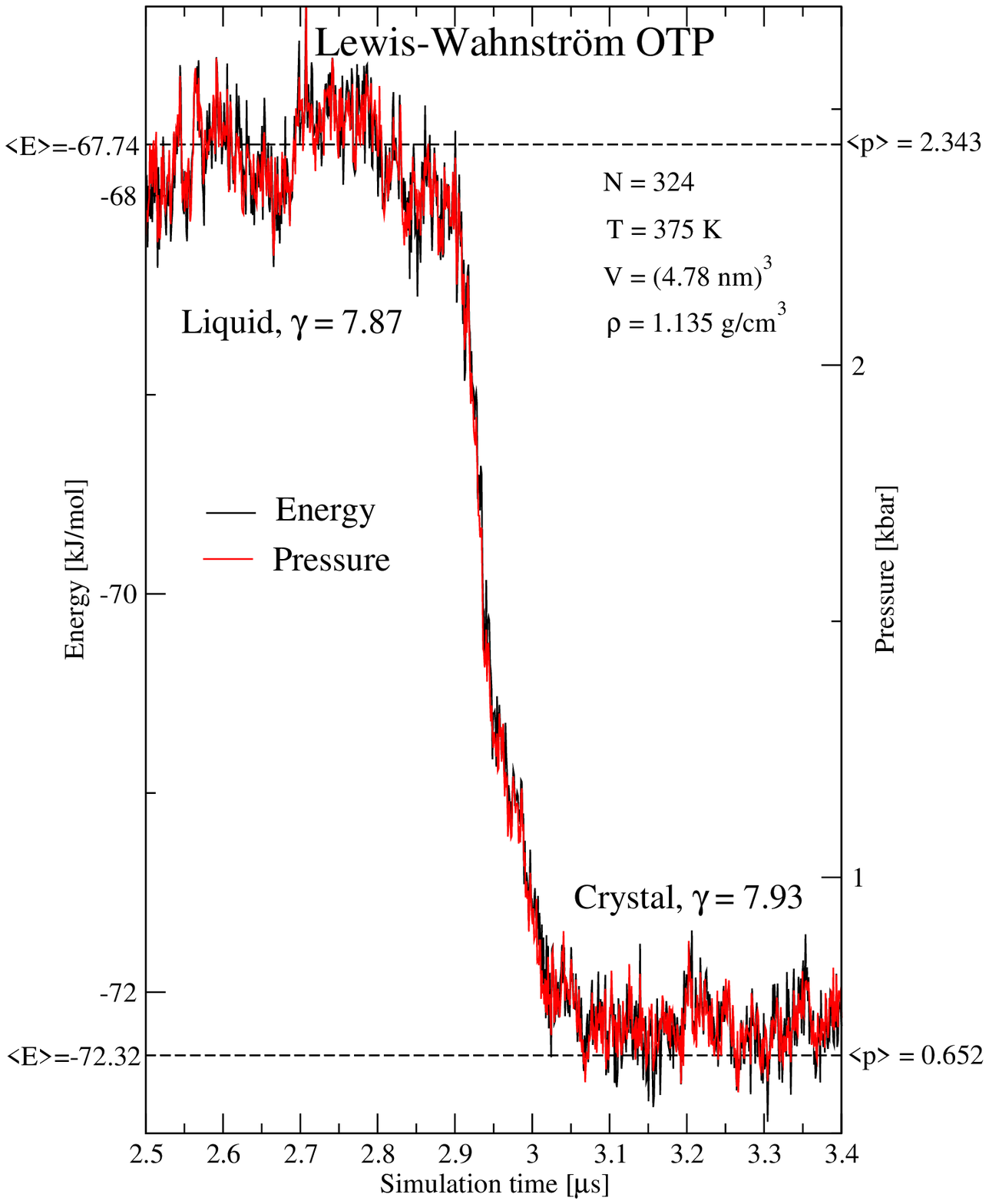} 
 \onefigure[width=4.5cm]{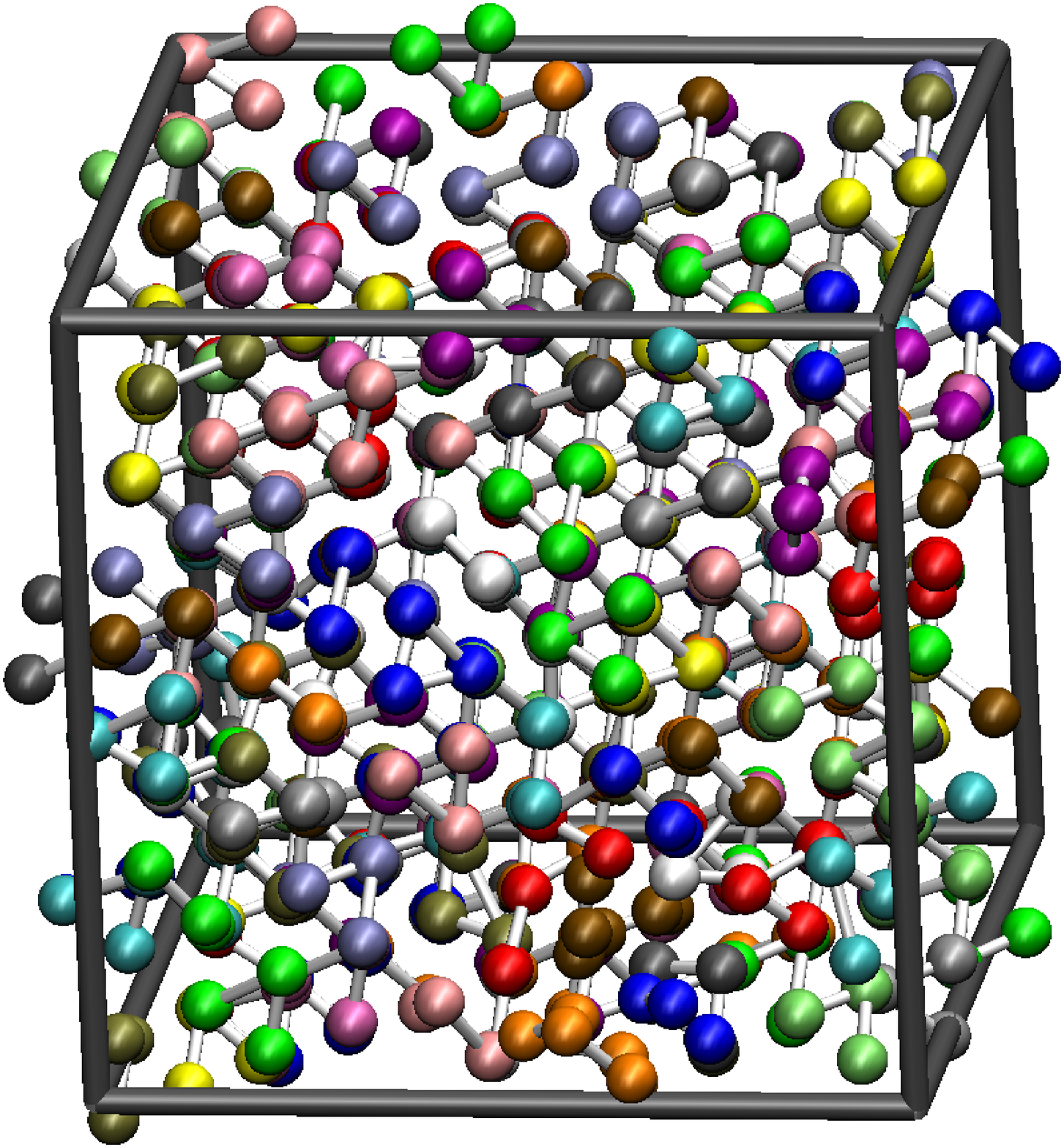}
 \onefigure[width=7.5cm]{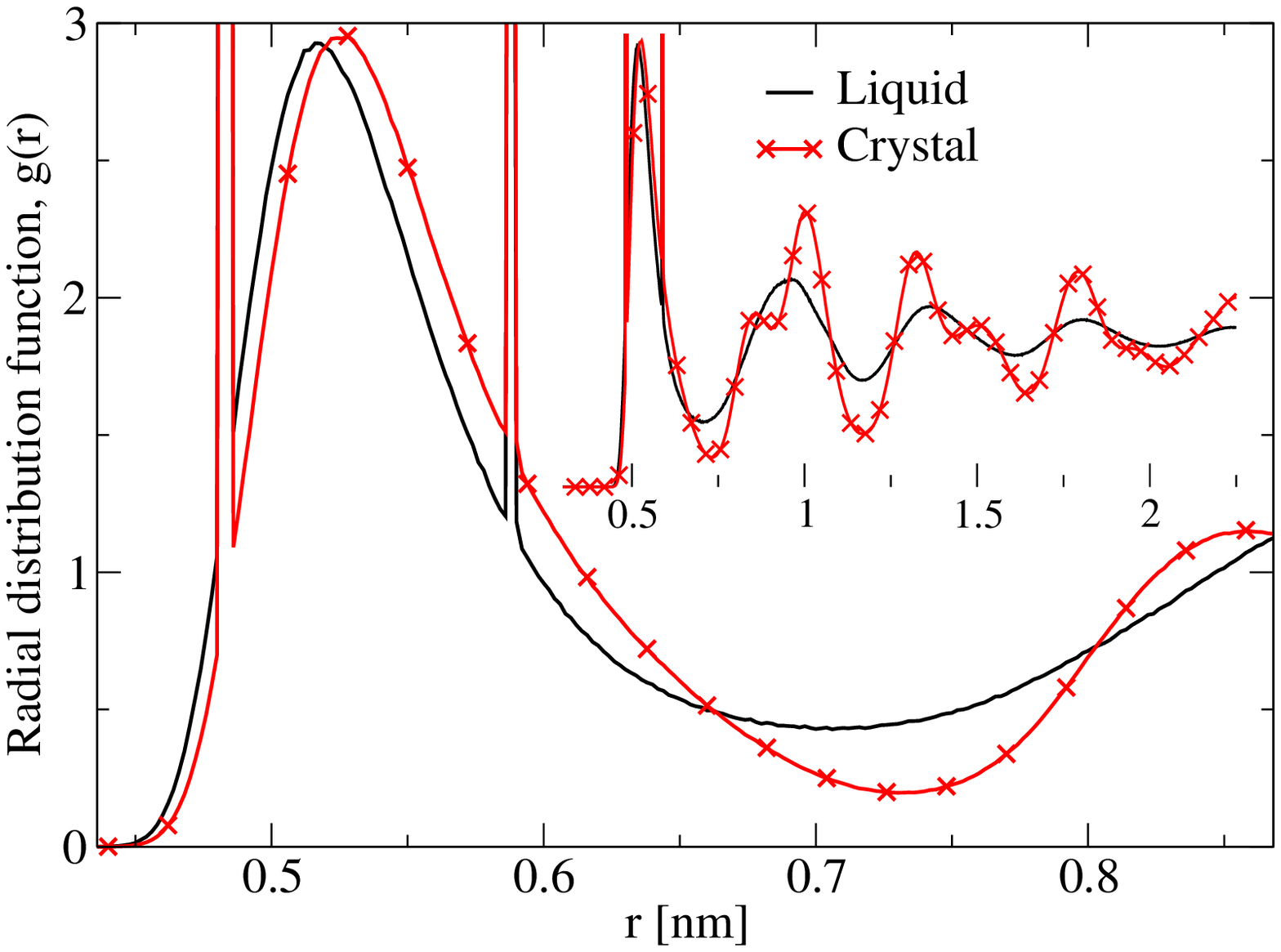}
\end{center}
\caption{Crystallization of the supercooled Lewis-Wahnstr{\"o}m ortho-terphenyl (OTP) liquid where each molecule consists of three Lennard-Jones spheres with fixed bond lenghts and angles \cite{lew94}.
(a) Pressure (right) and energy (left) monitored as functions of time during crystallization at constant volume (T= 375 K). Both quantities were averaged over 1 ns; in this way the pressure /energy fluctuations directly reflect the virial / potential energy fluctuations. The horizontal dashed lines give the liquid (upper line) and crystal (lower line), the averages of which were obtained from the simulation by averaging over times 0-2 $\mu$s and 5-10 $\mu$s, respectively. Both liquid and crystal show strong correlations, and the correlations are also present during the crystallization.
(b) The resulting OTP crystal. The crystal is an fcc-like crystal where the LJ atoms form a slightly distorted fcc lattice with random bond directions.
(c) Radial distribution functions of liquid and crystalline phases. The two spikes present in both phases come from the fixed bond lenghts.
}
\label{Ulf_figures}
\end{figure*}
A different far-out-of-equilibrium situation is that of crystallization of a supercooled liquid monitored at fixed volume and temperature. To the best of our knowledge crystallization of the LW OTP model has not been reported before, but Fig. \ref{Ulf_figures} shows that for simulations over microseconds the supercooled liquid crystallizes at T=375 K and $\rho=1.135 {\rm g/cm^3}$. The crystal (a distorted fcc structure with random bond orientations) is shown in Fig. \ref{Ulf_figures}(b). Figure \ref{Ulf_figures}(a) shows how time-averaged pressure and energy develop during crystallization. Contributions to pressure and energy from momenta are virtually constant after averaging over 1 ns, so strong $WU$ correlations manifest themselves in strong averaged-pressure /averaged-energy correlations. Clearly averaged pressure and averaged energy follow each other closely also during crystallization. This confirms the above finding, as well as those of Ref. \cite{bai08a} that strong correlations apply also for the crystalline phase of a strongly correlating liquid. Note that the slope $\gamma$ is virtually unaffected by the crystallization. The persistence of strong pressure-energy correlations during crystallization and the insignificant change of the $\gamma$'s are remarkable, because physical characteristics are rarely unaffected by a first-order phase transition. This shows, again, that the property of strong pressure-energy correlations pertains to the intermolecular potential, not to the particular configurations under study.

Figure \ref{Ulf_figures}(c) shows the radial distribution functions for the liquid and crystalline phases.

\section{Concluding remarks}
\begin{figure*}
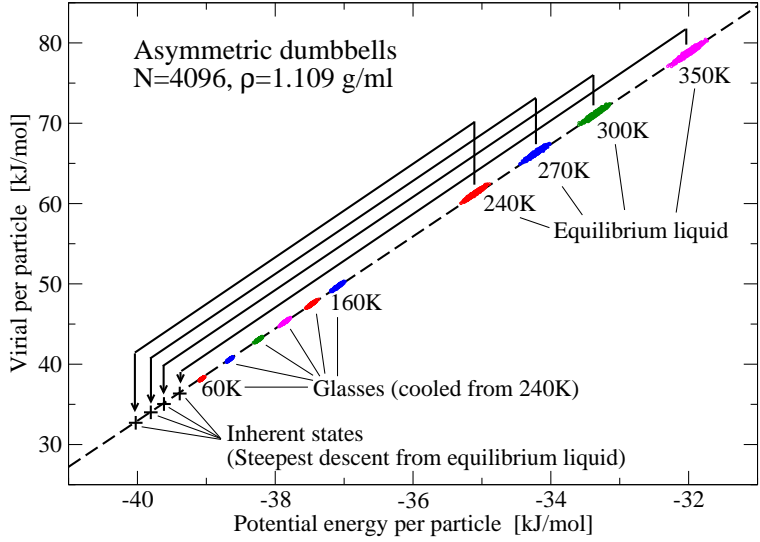

\begin{center}
 \onefigure[width=10.0cm]{fig4.eps} 
\end{center}
\caption{$WU$ plot for the asymmetric dumbbell model. The upper right corner shows data for simultaneous values of virial and potential energy for four equilibrium simulations (T=240-350 K). When quenching each of these to zero temperature in order to identify the inherent states, the crosses are arrived at. The intermediate points are glasses prepared by different cooling rates, i.e., out-of-equilibrium systems generated by cooling in 1 ns from 240 K to the temperature in question. This plot shows that strong virial / potential energy correlations are not limited to thermal equilibrium situations.
}
\label{WvsEpotLiq_Glass_Inh}
\end{figure*}

The above out-of-equilibrium simulations show that the property of strong $WU$ correlation is not confined to thermal equilibrium. Strongly correlating liquids have a particularly simple configuration space. These results have significance for any out-of-equilibrium situation. Consider the potential energy landscape picture of viscous liquid dynamics \cite{landscape} according to which each configuration has an underlying inherent state defined via a deepest-descent quench, a state that contains most information relevant to the slow dynamics (that may be regarded as jumps between different inherent states \cite{landscape}). Figure \ref{WvsEpotLiq_Glass_Inh} shows a $WU$ plot of the asymmetric dumbbell model in different situations: equilibrium states (upper right) and their corresponding inherent states (lower left, one quench per temperature), and glasses at different temperatures in between. A glass is an out-of-equilibrium state, and inherent states may be regarded as zero-temperature glasses. Altogether this plot shows once again that strong correlations are present also far from equilibrium

If $W_{\rm IS}$ is the inherent state virial, Eq. (\ref{Wdef}) may be rewritten as follows

\begin{equation}\label{WISdef}
p \,=\,
\frac{W_{\rm IS}}{V}+\frac{W-W_{\rm IS}}{V}+\frac{Nk_BT}{V}
\,.
\end{equation}
The fact that strong pressure-energy correlations apply also out of equilibrium validates the argument presented in Ref. \cite{bai08b} relating to the beautiful 2002 paper by Mossa, La Nave, Sciortino, and Tartaglia \cite{mos02}, who simulated four different aging scenarious of LW OTP (temperature jumps at constant volume, pressure jumps at constant temperature, pressure jumps at constant temperature in the glass phase, and isobaric heatings of a glass). Mossa {\it et al.} showed that all their findings may be rationalized in the following equation of state for the averge pressure involving the average inherent energy, $e_{\rm IS}$:

\begin{equation}\label{mossa}
p(T,V,\langle e_{\rm IS}\rangle)\,=\,
p_{\rm IS}(V,\langle e_{\rm IS}\rangle)+p_{\rm vib}(T,V,\langle e_{\rm IS}\rangle)\,.
\end{equation}
Thus for given temperature and volume knowledge of $\langle e_{\rm IS}\rangle$ in an out-of-equilibrium situation is enough to determine the average pressure (we follow the original notation, but $e_{\rm IS}=U_{\rm IS}$). As noted in Ref. \cite{bai08b} Eq. (\ref{mossa}) follows from Eq. (\ref{WISdef}): The $p_{\rm IS}$ term derives from the $W_{\rm IS}/V$ term of Eq. (\ref{WISdef}) because $\langle W_{\rm IS}\rangle$ is a (linear) function of $\langle e_{\rm IS}\rangle$. For any given state $W-W_{\rm IS}\cong \gamma (U-U_{\rm IS})=\gamma (e-e_{\rm IS})$ which -- assuming thermal equilibrium for the fast (vibrational) degrees of freedom -- upon averaging implies a contribution to the average pressure that depends on temperature and volume, with only a weak contribution from $\langle e_{\rm IS}\rangle$. In conjunction with the $Nk_BT/V$ term of Eq. (\ref{WISdef}) this gives the $p_{\rm vib}$ term of the Mossa {\it et al.} out-of-equilibrium equation of state. This derivation, incidentally, explains the subsequent finding by Sciortino and coworkers that Eq. (\ref{mossa}) does not apply for water models \cite{gio03} -- as we have seen water is not strongly correlating.

In conclusion, strong pressure-energy correlations reside in configuration space. This is consistent with the explanation provided in Ref. \cite{bai08b} that the strong correlations reflect the fact that an inverse power-law is a good approximation to the potential {\it at fixed-volume conditions}. In particular, the inverse power-law approximation yields almost correct Boltzmann probability factors \cite{sch08b}. The inverse power-law approximation is not exact, of course, but it reflects a hitherto overlooked approximately scale-invariance of the physics of a  large class of liquids \cite{sch08b} (van der Waals liquids and, possibly, some or all metallic liquids). Work is under way to further investigate the implications of this approximate scale invariance.

\acknowledgments
The centre for viscous liquid dynamics ``Glass and Time'' is sponsored by the Danish National Research Foundation (DNRF).


\begin{thebibliography}{99}
\bibitem{dumbbell_simulations} A system consisting of 512 asymmetric dumbbell molecules modelled as two Lennard-Jones spheres connected by a rigid bond was simulated. The dumbbells were parameterized to mimic toluene. A large sphere (mimicking a phenyl group) was taken from the Lewis-Wahnstr{\"o}m OTP model \cite{lew94} with the parameters $m_p=77.106$ u, $\sigma_p=0.4963$ nm and $\epsilon_p=5.726$ kJ/mol. A small sphere (mimicking a methyl group) was taken from UA-OPLS having $m_m=15.035$ u, $\sigma_m=0.3910$ nm and $\epsilon_m=0.66944$ kJ/mol. The bonds were kept rigid with a bond length of $d=0.29$ nm. The volume was $V=77.27$ nm$^3$ giving an average pressure of approximately 1 atm. The temperature was held constant at $T=130$ K using the Nos\'{e}-Hoover thermostat.
\end{thebibliography}
\end{document}